\documentclass[12pt]{iopart}

 \newcommand{\nn}{\nonumber\\} 
  
 \newcommand{\fk}[1]{\mbox{\boldmath$\scriptstyle #1$}}
 \newcommand{\f}[1]{\mbox{\boldmath$#1$}}
 \newcommand{\vau}{\mbox{\boldmath$v$}}
 \newcommand{\na}{\mbox{\boldmath$\nabla$}}
 \newcommand{\bea}{\begin{eqnarray}}
 \newcommand{\ea}{\end{eqnarray}}
 \newcommand{\eea}{\end{eqnarray}}
 \newcommand{\ord}{{\cal O}}

\begin{document}

\title[Aspects of Cosmic Inflation in Expanding Bose-Einstein Condensates]
{Aspects of Cosmic Inflation in Expanding Bose-Einstein Condensates}

\author{Michael Uhlmann, Yan Xu, and Ralf Sch\"utzhold\footnote[7]{To
whom correspondence should be addressed (schuetz@theory.phy.tu-dresden.de)}} 

\address{Institut f\"ur Theoretische Physik,
        Technische Universit\"at Dresden,
        01062 Dresden, Germany}

\begin{abstract}
Phonons in expanding Bose-Einstein condensates with wavelengths much 
larger than the healing length behave in the same way as quantum fields 
within a universe undergoing an accelerated expansion.
This analogy facilitates the application of many tools and concepts known 
from general relativity (such as horizons) and the prediction of the 
corresponding effects such as the freezing of modes after horizon crossing 
and the associated amplification of quantum fluctuations.
Basically the same amplification mechanism is (according to our standard 
model of cosmology) supposed to be responsible for the generation of the
initial inhomogeneities -- and hence the seeds for the formation of 
structures such as our galaxy -- during cosmic inflation 
(i.e., a very early epoch in the evolution of our universe).
After a general discussion of the analogy (\emph{analogue cosmology}), 
we calculate the frozen and amplified density-density fluctuations for 
quasi-two dimensional (Q2D) and three dimensional (3D) condensates which 
undergo a free expansion after switching off the (longitudinal) trap. 
\end{abstract}



\maketitle

\section{Introduction}

A quarter of a century ago, W.~G.~Unruh \cite{unruh} noticed an intriguing
analogy between (at a first glance) very different systems:  
Phonons in irrotational fluids behave in the same way as (quantum)
fields in curved space-times whose geometry is determined by the
effective metric and depends on the velocity of the fluid flow and the 
speed of sound etc.
This analogy facilitates a know-how transfer in both directions:
On the one hand, it allows us to apply all the concepts, tools, and
effects known from general relativity -- such as horizons -- to the
propagation of phonons (or other quasi-particles) in fluids and hence
leads to a better understanding of condensed-matter phenomena in terms
of universal geometrical concepts.  
On the other hand, this analogy opens up the opportunity of a
theoretical and possibly experimental investigation of exotic quantum 
effects known from cosmology such as Hawking radiation \cite{hawking}.

Unfortunately, it turns out that an experimental verification of the
Hawking effect by means of such an analogue system 
(i.e., a black hole analogue) is very difficult \cite{pessi}.
In this article, we shall focus on another exotic quantum phenomenon
known from cosmology which also involves the amplification of the
initial quantum vacuum fluctuations due to the presence of a horizon:
Within our standard model of cosmology, the early universe was nearly
homogeneous and basically all inhomogeneities 
-- including the seeds for the formation of large-scales structures
such as our galaxy -- originate from the quantum vacuum fluctuations
of a scalar field, which is called the inflaton.
During inflation (\cite{inflation}), which is a (conjectured) period
of accelerated expansion of the very early universe, these quantum
fluctuations were amplified due to the presence of a cosmic horizon
(cf.~Section~\ref{Horizon Analogues}).
These amplified fluctuations left an imprint in the cosmic microwave
background radiation -- small perturbations upon a homogeneous
background as measured by the WMAP satellite \cite{WMAP}.

Owing to the aforementioned analogy, the very same amplification
mechanism should also occur in appropriate fluids.
As an example, we shall consider the in some sense best understood
superfluids -- dilute atomic/molecular Bose-Einstein condensates
(BECs), see, e.g., \cite{Dalfovo} for review.
Apart from the good theoretical understanding, further advantages of
BECs are the various possibilities to manipulate 
(e.g., changing the speed of sound) and control them experimentally.
As it will become evident later, the free expansion of a BEC after
switching off the trap -- which is a standard procedure in
time-of-flight experiments/measurements, for example -- leads to the
formation of an effective cosmic horizon and, consequently, to the
amplification of the initial quantum fluctuations of the phonon
modes.
These amplified fluctuations manifest themselves in potentially
measurable small-scale density (and phase) variations.

In order to study this effect for rather general conditions, we shall
consider BECs with non-specified power-law self-interactions in an
arbitrary number of (spatial) dimensions.
After a discussion of the effectively lower-dimensional behaviour of
strongly constrained BECs in Section~\ref{Dimensional Reduction}, 
the effective metric governing the propagation of phonons is derived
in Section~\ref{Effective Geometry} for arbitrary power-law
self-interactions in any dimension.
Section~\ref{Co-Moving Coordinates and Scaling} is devoted to the
introduction of co-moving coordinates (in complete analogy to
cosmology) and to the scaling behaviour of the expanding condensate.
The analogy to cosmology is further elaborated in 
Section~\ref{Horizon Analogues} by applying the concept of an
effective horizon, whose existence generates the aforementioned
amplification mechanism.
The spectrum and magnitude of the resulting density fluctuations are
calculated in Sections~\ref{2D} and~\ref{3D} for the quasi-two and the  
three-dimensional case, respectively, assuming the usual quartic
coupling. 

\section{Dimensional Reduction}\label{Dimensional Reduction}

Since we shall consider an arbitrary number of spatial dimensions later
on, let us first discuss the behaviour of strongly confined and hence
effectively lower-dimensional Bose-Einstein condensates.
In three spatial dimensions, Bose-Einstein condensates are described
by the Lagrange density (${\hbar=1}$ throughout) \cite{Kamchatnov}
\bea
\label{L-original}
{\cal L} 
= 
\frac{i}{2} 
(\Psi^*\dot\Psi - \dot\Psi^*\Psi) - \frac{|\na\Psi|^2}{2m}
- V_{\rm ext}(\f{r},t)|\Psi|^2 - \frac{g}{2}\,|\Psi|^4 
\,,
\ea
with $m$ being the mass of the bosons, $V_{\rm ext}$ the external and
generally space-time dependent one-particle trapping potential, and
the two-particle coupling $g$ in $s$-wave approximation, which is
related to the $s$-wave scattering length $a_s$ via 
${g=4\pi a_s/m}$.
If the external one-particle trapping potential can be split up into 
a parallel and a static transversal part via 
\bea
\label{external}
V_{\rm ext}(\f{r},t)=
V^\|_{\rm ext}(\f{r}_\|,t)+V^\perp_{\rm ext}(\f{r}_\perp)
\,,
\ea
it is useful to decompose the order parameter $\Psi$ 
(and its quantum fluctuations ${\delta\hat\Psi}$) into a complete set of
real and time-independent functions ${\phi_\alpha(\f{r}_\perp)}$ 
governing the transversal dependence (cf. \cite{Zaremba} where the density
of an elongated BEC is decomposed in a similar manner)
\bea
\label{decompose}
\Psi(\f{r},t)=
\sum_\alpha\phi_\alpha(\f{r}_\perp)\psi_\alpha(\f{r}_\|,t)
\,.
\ea
The time evolution of the coefficient functions 
${\psi_\alpha(\f{r}_\|,t)}$ is determined by the reduced Lagrangian
density 
\bea
\label{L-reduced}
{\cal L}_\|
&=&
\int dV^{3-D}_\perp\,{\cal L} 
\nn
&=& 
\sum_{\alpha\beta}\left[\frac{i}{2}
(\psi^*_\alpha\dot\psi_\beta-\dot\psi^*_\alpha\psi_\beta)
-\frac{1}{2m}(\na\psi^*_\alpha)\cdot(\na\psi_\beta)
-V^\|_{\rm ext}\psi^*_\alpha\psi_\beta
\right]
\int dV^{3-D}_\perp\,\phi_\alpha\phi_\beta
\nn
&&-\sum_{\alpha\beta}\psi^*_\alpha\psi_\beta
\int dV^{3-D}_\perp\,\phi_\alpha
\left(-\frac{\na_\perp^2}{2m}+V^\perp_{\rm ext}\right)\phi_\beta
\nn
&&-
\frac{g}{2}
\sum_{\alpha\beta\gamma\delta}
\psi^*_\alpha\psi_\beta\psi^*_\gamma\psi_\delta
\int dV^{3-D}_\perp\,
\phi^*_\alpha\phi_\beta\phi^*_\gamma\phi_\delta
\,.
\ea
If we now choose the functions ${\phi_\alpha(\f{r}_\perp)}$ to be
the orthonormal eigenfunctions of the self-adjoint operator $\cal K$ 
with eigenvalues $\Omega_\alpha$
\bea
\label{eigenvalues}
{\cal K}\phi_\alpha
=
\left(-\frac{\na_\perp^2}{2m}+V^\perp_{\rm ext}\right)\phi_\alpha
=
\Omega_\alpha\phi_\alpha
\,,
\ea
the reduced Lagrangian density simplifies to
\bea
\label{L-simple}
{\cal L}_\|
&=&
\sum_\alpha\left[
\frac{i}{2}(\psi^*_\alpha\dot\psi_\alpha-\dot\psi^*_\alpha\psi_\alpha)
-\frac{1}{2m}|\na\psi_\alpha|^2
-(V^\|_{\rm ext}+\Omega_\alpha)|\psi_\alpha|^2
\right]
\nn
&&-
\frac{g}{2}
\sum_{\alpha\beta\gamma\delta}
\psi^*_\alpha\psi_\beta\psi^*_\gamma\psi_\delta
\int dV^{3-D}_\perp\,
\phi^*_\alpha\phi_\beta\phi^*_\gamma\phi_\delta
\,.
\ea
The last term induces a coupling of different modes, which complicates
the solution. 
However, if we assume that the coupling term is sufficiently small and
the lowest mode ${\alpha=0}$ dominates ${|\psi_0|\gg|\psi_{\alpha>0}|}$, 
we may estimate the population of the higher modes in analogy to
stationary perturbation theory:
The mixing between the lowest mode ${\alpha=0}$ and all higher modes 
${\alpha>0}$ is small if the energy differences 
${\Delta\Omega_\alpha=\Omega_\alpha-\Omega_0}$ are large compared to
the transition matrix elements of the interaction Hamiltonian, i.e.,  
\bea
\label{transition}
\Delta\Omega_\alpha
\gg 
g|\psi_0|^2\int dV^{3-D}_\perp\,
\phi^*_\alpha\phi_\beta\phi^*_\gamma\phi_\delta
\,.
\ea
Since the modes $\phi_\alpha$ are normalised 
${\phi_\alpha\sim1/\sqrt{V^{3-D}_\perp}}$, this condition can be 
re-expressed in terms of the size of the transversal dimension
$a_\perp$ and the healing length ${\xi=1/\sqrt{g|\psi_0|^2 m}}$.
Therefore, imposing the conditions
\bea
\label{conditions}
\xi \gg a_\perp \gg a_s
\,,
\ea
where the latter requirement ${a_\perp \gg a_s}$ is necessary for the 
Gross-Pitaevski\v\i\/ Lagrangian density in Eq.~(\ref{L-original}) to
be valid, we arrive at the reduced Lagrangian density for the lowest
mode ${\psi=\psi_0}$
\bea
\label{L-final}
{\cal L}_\|
=
\frac{i}{2}(\psi^*\dot\psi-\dot\psi^*\psi)
-\frac{1}{2m}|\na\psi|^2
-V^\|_{\rm ext}|\psi|^2
-
\frac{g_\|}{2}\,|\psi|^4
\,,
\ea
where $\Omega_0$ has been absorbed into $V^\|_{\rm ext}$ and  
\bea
\label{g-reduced}
g_\|
=
g\int dV^{3-D}_\perp\,|\phi_0|^4
=
\frac{4\pi a_s}{m}\int dV^{3-D}_\perp\,|\phi_0|^4
\propto
\frac{a_s}{a_\perp^{3-D}}
\,,
\ea
denotes the reduced (lower-dimensional) coupling constant.
For example, the eigenfunctions 
${\phi_\alpha(\f{r}_\perp)=\phi_\alpha(z)}$ of a harmonic potential 
${V_\perp(z)=m\omega_z^2z^2/2}$ with the transversal length scale 
${a_\perp=1/\sqrt{m\omega_z}}$ are just the Legendre polynomials with
an equidistant energy spectrum.
In this case, the reduced coupling is given by 
${g_\|=g \sqrt{m \omega_z/2\pi}}$. 
However, it should be emphasised that the method presented above is
applicable to more general potentials satisfying the aforementioned
conditions as well. 

Note that if we relaxed condition~(\ref{conditions}) and the transverse
size of the condensate, $a_z$, would be of comparable to the scattering
length, $a_s/a_z = \mathcal O(1)$, we would obtain a lower-dimensional
condensate with different interactions due to density-dependent
corrections to the coupling coefficient
\cite{Petrov,Petrov_2,Kolomeisky,Lee}. 

\section{Effective Geometry}\label{Effective Geometry}

After having discussed the dimensional reduction, let us start with
the Lagrange density in an arbitrary number of spatial dimensions $D$
(where we omit the superscript $\|$ for the sake of conciseness).
In addition, we shall assume a more general self-coupling term
${|\psi|^{2N}}$ which will be motivated later 
\bea
\label{L-general}
{\cal L}
=
\frac{i}{2}(\psi^*\dot\psi-\dot\psi^*\psi)
-\frac{1}{2m}|\na\psi|^2
-V_{\rm ext}|\psi|^2
-
\frac{g}{2}\,|\psi|^{2N}
\,.
\ea
Inserting the Madelung representation 
\bea
\label{Madelung}
\psi=\sqrt{\varrho}\,e^{iS}
\,,
\ea
the Lagrange density reads 
\bea
\label{L-Madelung}
{\cal L}
=  
-
\varrho\,\partial_t S
-
\frac{\varrho}{2m}(\na S)^2
-
\frac{(\na\sqrt{\varrho})^2}{2m}
-
V_{\rm ext}\,\varrho 
-\frac{g}{2}\,\varrho^N
\,.
\ea
As usual, variation w.r.t.~$S$ yields the equation of continuity and
w.r.t.~$\varrho$ the Bernoulli equation with the mean-field velocity
$\vau=\na S/m$ and the specific pressure
${p(\varrho,\na^2\sqrt{\varrho})}$. 
Assuming that the density profile is sufficiently smooth, i.e., that
the relevant length scales are much larger than the healing length 
(Thomas-Fermi approximation), we neglect the quantum pressure term  
${(\na\sqrt{\varrho})^2}$.
Linearisation ${S=S_0+\delta S}$ and
${\varrho=\varrho_0+\delta\varrho}$ around a background solution $S_0$
and $\varrho_0$ yields the second-order Lagrange density
(cf.~\cite{Fischer_2,Fedichev_2,Fischer,Stone})
\bea
\label{L-2}
{\cal L}^{(2)}
=  
-
\delta\varrho\,\partial_t\delta S
-
\frac{\varrho_0}{2m}(\na\delta S)^2
-
\delta\varrho\,\vau_0\cdot\na\delta S
-
\frac{g_N}{2}\,\delta\varrho^2
\,,
\ea
where we have introduced the effective coupling
\bea
\label{g-eff}
g_N=g\,\frac{N(N-1)}{2}\,\varrho_0^{N-2}
\,.
\ea
For the usual case $N=2$, we have $g_N=g$, but one should still bear
in mind that we are considering an arbitrary number of spatial
dimensions $D$. 
Note that there are no sound waves at all for $N=1$, since the theory 
is non-interacting in that case.

In the Thomas-Fermi approximation, the linearised Bernoulli equation
can be solved for the density fluctuations 
\bea
\label{density}
\delta\varrho=-\frac{\partial_t+\vau_0\cdot\na}{g_N}\,\delta S
\,,
\ea
and inserting this result back into Eq.~(\ref{L-2}), we obtain the
effective Lagrangian for the phase fluctuations ${\phi=\delta S}$ only  
\bea
\label{L-phase}
{\cal L}^{(2)}_{\rm eff}
=
\frac{1}{2 g_N}\left(\dot\phi + \vau_0\cdot\na\phi\right)^2
-\frac{\varrho_0}{2m}(\na\phi)^2
\,.
\ea
Even in an arbitrary number of spatial dimensions $D$ 
(the difficulties for $D=1$ will be discussed below) and for 
general self-coupling $N>1$, this Lagrangian is completely equivalent
to that of a free (minimally coupled) scalar field in a curved
space-time (e.g., \cite{Birrell})
\bea
\label{L-phi}
{\cal L}^{(2)}_{\rm eff}
=
\frac12\sqrt{|\mathrm{g}_{\rm eff}|}\,
(\partial_\mu \phi) \mathrm{g}^{\mu\nu}_{\rm eff} (\partial_\nu \phi)
\,,
\ea
provided that we insert the 
Painlev\'e-Gullstrand-Lema\^itre (PGL) metric \cite{PGL}
\bea
\label{PGL}
\mathrm{g}_{\mu\nu}^{\rm eff}
&=&
A_D^{(N)}
\left( 
\begin{array}{cr} 
c^2_N-{\vau}^2_0 \, &\, {\vau}_0 \\ 
{\vau}_0\, & -{\f{1}} 
\end{array} 
\right)
\,,
\nn
\mathrm{g}^{\mu\nu}_{\rm eff}
&=&
\frac{1}{A_D^{(N)}c_N^2}
\left( 
\begin{array}{cc} 
1 \, &\, {\vau}_0 \\
{\vau}_0 \, &\, {\vau}_0\otimes{\vau}_0-c^2_N\f{1}
\end{array} 
\right)
\,,
\ea
with the speed of sound ${c_N^2=g_N\varrho_0/m}$ and the conformal
factor 
\bea
\label{conformal}
A_D^{(N)}=\left(\frac{c_N}{g_N}\right)^{2/(D-1)}
\,.
\ea
This expression already indicates problems in one spatial dimension
$D=1$ due to conformal invariance of the scalar field action
(\ref{L-phi}) in 1+1 dimensions.
Without the introduction of an additional dilaton field, the
identification of an effective metric is only possible if 
${c_N/g_N=\rm const}$, for example if ${g=\rm const}$ and $N=3$, 
see also the next Section.

\section{Co-Moving Coordinates and Scaling}
\label{Co-Moving Coordinates and Scaling}

Since phase fluctuations in Bose-Einstein condensates behave 
(in the Thomas-Fermi approximation) exactly as a scalar field in a
specific curved space-time described by the effective metric in
Eq.~(\ref{PGL}), it will be useful to investigate this metric with the
tool known from general relativity. 
The effective line element reads 
\bea
\label{line}
ds^2_{\rm eff}=A_D^{(N)}
\left([c^2_N-{\vau}^2_0]dt^2+2\vau_0\cdot d\f{r}dt-d\f{r}^2\right)
\,.
\ea
In analogy to cosmology, we shall assume local isotropy and
homogeneity -- which will be a good approximation in the centre of the
BEC cloud.
Since the coupling $g$ is supposed to be constant for simplicity, this
assumption implies a spatially homogeneous but possibly time-dependent
density ${\varrho_0=\varrho_0(t)}$ and effective coupling
${g_N=g_N(t)}$.  
Furthermore, after a suitable re-definition of the origin of our
coordinate system, we may set $\vau_0\propto\f{r}$ due to the 
presumed local isotropy and homogeneity.
Insertion of this {\em ansatz} into the equation of continuity yields
(\cite{Fedichev_2,Fischer})
\bea
\label{ansatz}
\varrho_0(t)=\frac{\varrho_0(t=0)}{b^D(t)}
\;\leftrightarrow\;
\vau_0(t,\f{r})=\frac{\dot b}{b}\,\f{r}
\,,
\ea
which allows us to describe $\varrho_0(t)$ and $\vau_0(t,\f{r})$ as
well as $g_N(t)$ by means of a single time-dependent scaling parameter  
$b(t)$ satisfying the initial condition $b(t=0)=1$ and $\dot b(t=0)=0$. 
As we know from general relativity, an off-diagonal metric such as in
Eq.~(\ref{line}) can be diagonalised by introducing co-moving spatial 
coordinates via (cf. the scaling transformation for BECs in
\cite{Castin,Kagan})
\bea
\label{co-moving}
\f{\rho}=\frac{\f{r}}{b(t)}
\,\to\;
d\f{\rho}=\frac{d\f{r}-\vau_0dt}{b(t)}
\,\to\;
ds^2_{\rm eff}=A_D^{(N)}
\left(c^2_Ndt^2-b^2d\f{\rho}^2\right)
\,.
\ea
In addition, we may transform from the laboratory time $t$ to the
effective proper (co-moving wrist-watch) time $\tau$ in order to
eliminate the factor in front of $d\tau^2$
\bea
\label{proper}
\tau=\int dt\,\sqrt{A_D^{(N)}}\,c_N
\,\to\;
ds^2_{\rm eff}=d\tau^2-A_D^{(N)}b^2d\f{\rho}^2
\,,
\ea
arriving at the standard Friedmann-Robertson-Walker representation,
see, e.g., \cite{Birrell}.
Let us investigate the remaining factor ${A_D^{(N)}b^2}$ a bit further: 
Since ${\varrho_0(t)\propto b^{-D}(t)}$ according to
Eq.~(\ref{ansatz}), we obtain ${g_N(t)\propto b^{-D(N-2)}(t)}$ 
from Eq.~(\ref{g-eff}) as well as ${c_N(t)\propto b^{-D(N-1)/2}(t)}$,
i.e., the factor in front of ${d\f{\rho}^2}$ scales as 
\bea
\label{factor}
A_D^{(N)}b^2
\propto 
b^{2+D(N-3)/(D-1)}
\,,
\ea
according to Eq.~(\ref{conformal}).
Interestingly, the exponent vanishes for
\bea
\label{vanish}
N=\frac{2}{D}+1
\,,
\ea
e.g., for $D=2$ and $N=2$, or $D=3$ and $N=5/3$, or $D=1$ and $N=3$.
As we have observed above, special care is required for the latter
case $D=1$, but for $N=3$ we can indeed introduce an effective metric
and the conformal factor can be chosen at will since it does not
enter the calculation.

If the condition in Eq.~(\ref{vanish}) is satisfied and thus the
exponent in Eq.~(\ref{factor}) vanishes, the transformed metric in 
Eq.~(\ref{proper}) is flat and hence the wave equation for the phonon
modes becomes trivial in terms of the coordinates $\tau$ and
$\f{\rho}$ 
\bea
\label{trivial}
\left(\frac{\partial^2}{\partial\tau^2}-
\frac{1}{A_D^{(N)}(t=0)}\,\frac{\partial^2}{\partial\f{\rho}^2}
\right)
\phi=0
\,,
\ea
i.e., a solution $\phi(\tau,\f{\rho})$ is independent of the external
time dependence mediated via the scaling parameter $b(t)$.
It turns out that this perfect scaling is not only valid for the
phonon modes, but can be extended to the full field operator:
If we start from the equation of motion in the Heisenberg picture 
\bea
\label{full}
i\frac{\partial}{\partial t}\hat\Psi=
\left(-\frac{1}{2m}\,\frac{\partial^2}{\partial\f{r}^2}
+V_{\rm ext}(t,\f{r})+
g\,(\hat\Psi^\dagger)^{N-1}\hat\Psi^{N-1}\right)\hat\Psi
\,,
\ea
again with general $s$-wave coupling ($N$) and in $D$ spatial
dimensions, we may account for an arbitrarily time-dependent external
one-particle trapping potential ${V_{\rm ext}(t,\f{r})}$ as long as it
is purely harmonic at all times 
\bea
\label{harmonic}
V_{\rm ext}(t,\f{r})=\frac{m}{2}\,\omega^2_{\rm ext}(t)\f{r}^2
\,,
\ea
by inserting the scaling {\em ansatz} derived above (cf., \cite{Kagan})
\bea
\label{scaling}
\hat\Psi(t,\f{r})
=
\frac{\exp\{im\vau_0^2(t,\f{r})/2\}}{\sqrt{b^D(t)}}\,
\hat\psi(\tau,\f{\rho})
\,.
\ea
The numerator ${\exp\{im\vau_0^2(t,\f{r})/2\}}$ of the pre-factor
reproduces $\vau_0$ and the denominator $\sqrt{b^D(t)}$ accounts for
${\varrho_0(t)=\varrho_0(t=0)/b^D(t)}$ and ensures the correct
commutation relations for ${\hat\psi(\tau,\f{\rho})}$, i.e.,
\bea
\label{commutation}
[\hat\psi(\tau,\f{\rho}),\hat\psi^\dagger(\tau,\f{\rho'})]
=
\delta^D(\f{\rho}-\f{\rho'})
\,.
\ea
For a flat metric $N=1+2/D$, the proper time is 
(independently of $D$) determined as  
\bea
\label{flat}
\tau=\int\frac{dt}{b^2(t)}
\,,
\ea
and $\f{\rho}$ is of course still given by ${\f{\rho}=\f{r}/b(t)}$. 
Finally, if we arrange the scale parameter $b(t)$ according to 
\bea
\label{parameter}
\ddot b(t)+\omega^2_{\rm ext}(t)b(t)
=
\frac{\omega^2_{\rm ext}(t=0)}{b^3(t)}
=
\frac{\omega^2_0}{b^3(t)}
\,,
\ea
the equation of motion for ${\hat\psi(\tau,\f{\rho})}$ becomes
independent of $b(t)$ 
\bea
\label{perfect}
i\frac{\partial}{\partial\tau}\hat\psi(\tau,\f{\rho})=
\left(-\frac{1}{2m}\,\frac{\partial^2}{\partial\f{\rho}^2}
+
\frac{m}{2}\,\omega^2_0\f{\rho}^2
+
g\,(\hat\psi^\dagger)^{N-1}\hat\psi^{N-1}\right)\hat\psi(\tau,\f{\rho})
\,,
\ea
i.e., we obtain a perfect scaling solution of the full field operator 
exactly in those cases where the effective metric is flat. 
Note that the implications of this property of the full field operator
go far beyond the scaling of the hydrodynamic solution -- we also
obtain a perfect scaling of the quantum fluctuations to arbitrary
order and for large wavenumbers 
(provided that the $s$-wave approximation is still valid, of course), 
where hydrodynamic (Thomas-Fermi) solution breaks down.
It is not even necessary to assume the mean-field expansion, i.e., the
scaling also applies to non-condensed bosons. 

Interestingly, all three examples (from $D=1$ to $D=3$) are potentially
relevant for real physical systems:
\begin{itemize}
\item 
$D=1$ and $N=3$: In quasi-1D Bose-Einstein condensates, one may obtain
   an effective $|\psi|^6$ coupling if the perpendicular size $a_z$ is
   comparable to the $s$-wave scattering length $a_s$, see, e.g.,
   \cite{Petrov_2,Kolomeisky,Lee}.
\item 
$D=2$ and $N=2$: This is the usual quasi-2D Bose-Einstein condensate with
   quartic coupling, e.g., \cite{Dalfovo}.
\item 
$D=3$ and $N=5/3$: Even though the exponent $N=5/3$ may seem somewhat
   unnatural, it appears in an effective description of a weakly
   interacting two-component Fermi gas in the BCS state, where the
   equation of state is determined by the Fermi energy
   ${E_{\rm F}\propto\varrho^{5/3}}$, and which therefore also shows
   scaling behaviour, see, e.g., \cite{Fedichev_2}.
\end{itemize}
%

\section{Horizon Analogues}\label{Horizon Analogues}

Apart from the question of whether it can be cast into a flat
space-time form by means of an appropriate coordinate transformation
or not, the emergence of a non-trivial effective metric in
Eq.~(\ref{PGL}) suggests the application of concepts known from
general relativity, such as horizons \cite{inflation,Wald,Hawking}. 
Generally speaking, horizons correspond to a loss of causal
connectivity, i.e., events beyond a horizon have no influence or
cannot be influenced.
The most prominent example is the event horizon of a
(classical) black hole, beyond which everything is trapped forever 
(i.e., nothing can ever escape to infinity).

On the other hand, in cosmology, two slightly different horizon
concepts play a more important role due to the large-scale homogeneity
and isotropy -- the particle horizon and the apparent horizon. 
The particle horizon always refers to a chosen trajectory and
indicates the border to the space-time region which cannot be reached
by any signal starting from this trajectory or from where no signal
can reach this trajectory. 
The apparent horizon depends on the chosen coordinates and is
(roughly speaking) defined as the border beyond which all closed
two-surfaces can either only expand or only contract w.r.t.~the chosen 
coordinates. 

As it turns out, the concepts of the particle and the apparent horizon 
can be applied to expanding Bose-Einstein condensates.
Let us start with the particle horizon and choose the trajectory
${\f{r}=0\to\f{\rho}=0}$ for which the particle horizon can
conveniently be determined using the metric in Eq.~(\ref{co-moving}). 
The question is: which values of the co-moving coordinate $\f{\rho}$
can a sound wave reach by starting at the origin $\f{\rho}=0$ at time
$t$ and being described by a null line ${ds^2_{\rm eff}=0}$, i.e., 
how far can it travel?
According to Eq.~(\ref{co-moving}), this length can be calculated via
the following integral (cf. also \cite{inflation})
\bea
\label{particle-h}
\Delta\rho_{\rm horizon}(t)
=
\int\limits_t^\infty dt'\,\frac{c_N(t')}{b(t')}
=
c_N(t=0)\int\limits_t^\infty\frac{dt'}{b^{1+D(N-1)/2}(t')}
\,.
\ea
If the above integral converges to a finite result
${\Delta\rho_{\rm horizon}(t)}$, this corresponds to a particle
horizon since no point beyond this co-moving coordinate can be reached
by sound waves anymore.

For a free expansion of the condensate 
${V_{\rm ext}(t\uparrow\infty)=0}$, we obtain 
${b(t\uparrow\infty)\propto t}$ at late times according to 
Eq.~(\ref{parameter}) and thus a horizon exists for all $N>1$.
The occurrence of a horizon might be puzzling if the metric in terms of
the proper time $\tau$ in Eq.~(\ref{proper}) is flat -- but this
puzzle can be resolved by the observation that the proper time $\tau$
reaches a finite value ${\tau(t\uparrow\infty)<\infty}$ in the limit
of arbitrarily late laboratory times in that case, cf.~Eq.~(\ref{flat}).

In contrast to the particle horizon, which necessitates the knowledge
of the full future (or past), the apparent horizon can be identified
by means of the configuration at a certain instant of time only. 
Instead of a trajectory as for the particle horizon, we have to choose
a specific coordinate system (time slices) in order to define the
apparent horizon.
The coordinates ${(\tau,\f{\rho})}$ leading to a flat metric, 
for example, do of course not allow the introduction of an apparent
horizon. 
On the other hand, since we observe the expanding condensate using the
laboratory coordinate system ${(t,\f{r})}$, we shall choose these
coordinates instead.
For a spherically symmetric metric as in Eq.~(\ref{PGL}), the apparent
horizon is determined by ${g_{00}^{\rm eff}=0}$, i.e., where the
velocity of the condensate $\vau_0$ exceeds the speed of sound $c_N$ 
\bea
\label{apparent-h}
\vau_0^2(t,r_{\rm horizon})=c^2_N(t)
\,\to\;
r_{\rm horizon}(t)\propto\frac{b^{1-D(N-1)/2}}{\dot b}
\,.
\ea
The very intuitive interpretation is that no sound wave can enter the
region ${r<r_{\rm horizon}}$ from the outside. 

For perfect scaling $N=1+2/D$ and a freely expanding condensate, 
the apparent horizon settles down at late times 
(when $\dot b$ becomes constant) to a finite value 
${r_{\rm horizon}(t\uparrow\infty)=r_{\rm horizon}^\infty>0}$, otherwise
it may increase or decrease forever (depending on $D$ and $N$).
Note that both, the apparent and the particle horizon are never at
rest w.r.t.~the co-moving coordinate $\f{\rho}$ but always
decreasing. 
As a result, the wavelength of all sound modes
${\exp\{i\f{\kappa}\cdot\f{\rho}\}}$ exceeds the horizon size at some
point of time (horizon crossing) in view of the permanent stretching
of modes due to the expansion of the condensate. 
After the wavelength exceeds the horizon size, the regions of higher
and lower pressure cannot interact anymore and hence the modes stop
oscillating (freezing of modes).
As a result, comparing each mode with a harmonic oscillator, the
momentum and its variance decrease drastically -- and according to the
Heisenberg uncertainty relation, the complementary variance must
increase in order to compensate this.
Ergo, the initial ground/vacuum state gets squeezed, which amplifies
the quantum fluctuation in a certain direction. 
This admittedly rather intuitive picture 
(horizon crossing, freezing, and squeezing) applies to both, the
expanding universe (provided a horizon exists, such as during cosmic
inflation) as well as expansing Bose-Einstein condensates in a very
similar way and must of course be further supported by explicit
calculations, see the next Section. 

Note that analogue horizons 
(and perhaps horizons in real gravity as well) 
are only low-energy effective concepts since, for very small
wavelengths (below the healing length), the quantum pressure term
becomes important leading to group and phase velocities exceeding the
usual speed of sound.   
Furthermore, as the healing length increases during the expansion, 
the late-time limit in the derivation of the particle horizon is not 
strictly valid. 
Nevertheless, the main effects such as the amplification of quantum
fluctuations are not modified drastically, see the next Section. 

\section{Density-Density Correlations in quasi-2D}\label{2D}

According to the results of Section~\ref{Co-Moving Coordinates and Scaling},
the time-evolution of a quasi-two-dimensional condensate with the usual
quartic coupling (i.e., $N = D = 2$) is particularly simple to solve due
to its perfect scaling behaviour -- which enables us to derive its
dynamics exactly (e.g., including the quantum-pressure corrections).
If we assume that the initial trapping potential is harmonic, the free
expansion after switching off the longitudinal trap 
$V_\|^{\rm ext}(t>0)=0$ is fully described by the time-evolution of
the scaling parameter according to Eq.~(\ref{parameter})
\bea
\label{free-2D}
b(t)=\sqrt{1+\omega^2_0t^2}
\,.
\ea
The associated proper time $\tau$ can be calculated from Eq.~(\ref{flat})
\bea
\label{arctan}
\tau(t>0)=\frac{\arctan\omega_0 t}{\omega_0}
\,.
\ea
Since $\tau$ quickly approaches a finite value at late times
$t\uparrow\infty$, the fluctuations freeze in and assume the value at
proper time $\tau=\pi/(2\omega_0)$.
In view of the perfect scaling discussed in 
Sec.~\ref{Co-Moving Coordinates and Scaling}, both the background and
the phonon modes -- expressed in terms of $\f{\rho}$ and $\tau$ -- 
behave in the same way before and during the expansion, leaving the
relative correlation function
\bea
\label{correlation}
C(\f{\rho}, \f{\rho}') 
= 
\frac{
\langle 
\delta\hat\varrho(t,\f{\rho})\delta\hat\varrho(t,\f{\rho}')
\rangle
}{\varrho_0(t,\f{\rho}) \varrho_0(t,\f{\rho}')}
= 
\frac{
\langle 
\delta\hat\varrho(\f{\rho})\delta\hat\varrho(\f{\rho}')
\rangle
}{\varrho_0(\f{\rho}) \varrho_0(\f{\rho}')}
\ea
unchanged.
Consequently, the spectrum of the relative density contrast in
co-moving coordinates after horizon crossing is given by the initial
correlation spectrum.
Besides the quasi-2D condensate with usual quartic coupling ($N=D=2$), 
this also holds for all other cases where perfect scaling occurs.

As discussed in Section~\ref{Horizon Analogues}, the apparent horizon
depends on the chosen time-slicing, i.e., the set of coordinates.
(The particle horizon is independent of the set of coordinates, but
requires the knowledge of the whole future evolution.)
Co-moving coordinates, albeit advantageous to describe the evolution
of the modes, are not suitable for describing the appearance of an
apparent horizon.
On the other hand, all observations are performed in laboratory time
$t$ and coordinates $\f{r}$ with the metric~(\ref{PGL}) showing the
formation of an apparent horizon according to Eq.~(\ref{apparent-h}).
For $N = 1 + 2/D$, it follows
\bea
r_{\rm horizon}(t)
=
\frac{c_N(0)}{\dot b(t)}
=
c_N(0)\frac{\sqrt{1+\omega^2_0t^2}}{\omega^2_0t}
\,,
\ea
in the homogeneous case.
While the trapping potential is still turned on ($t < 0$), 
the (apparent) horizon size $r_{\rm horizon}$ is infinite.
Clearly, without any expansion, all points  are causally connected via 
sound waves.
After releasing the condensate, however, the horizon settles down very
quickly ($\omega_0 t \gg 1$) at a finite position 
(in laboratory coordinates) $c_N(0)/\omega_0$, given by the initial
speed of sound $c_N(0)$ and trapping frequency $\omega_0$.

Since the (apparent) horizon approaches a finte position in
laboratory coordinates ($t,\f{r}$) and every phonon mode with a given
wavelength $\lambda$ expands with the condensate cloud, each phonon
mode will cross the horizon eventually.
Equivalently, in terms of the co-moving coordinates, the horizon size
decreases for all times (as long as the condensate is expanding) and
the (co-moving) wavenumber $\f{\kappa}$ remains constant.
As mentioned before, the initial correlation function of a quasi-2D
condensate translates directly into the frozen density contrast after
horizon crossing. 

In order to derive the density-density correlation function
quantitatively, we quantise the phonon modes within the initial
condensate $\vau_0=0$ using the approximation 
(centre of the BEC cloud) of a constant background  
density $\varrho_0=\rm const$.
After a normal mode expansion, the Lagrange
function~(\ref{L-Madelung}) reads 
\bea
L=\int d^2r\,{\cal L}=
\frac{V_Q}{4}\sum_{\fk{k}}
\left[
\left(g_{\rm 2D}+\frac{\f{k}^2}{4m\varrho_0}\right)^{-1}
(\delta\dot S_{\fk{k}})^2
-\frac{\varrho_0\f{k}^2}{m}(\delta S_{\fk{k}})^2
\right]
\,,
\ea
where $V_Q$ denotes the quantisation volume.
Note that, in contrast to Eq.~(\ref{L-2}), we did not neglect the quantum
pressure term.
In a straightforward manner, the phase and density fluctuations can be
quantised via the introduction of creation 
($\hat a_{\fk{k}}^\dagger$) and annihilation operators 
($\hat a_{\fk{k}}$) which diagonalise the Hamiltonian.
The quantised phase and density fluctuations read
\bea
\delta\hat S_{\fk{k}}
&=&
\sqrt{\frac{1}{V_Q\omega_{\fk{k}}}
\left(g_{\rm 2D}+\frac{\f{k}^2}{4m\varrho_0}\right)}
\left(\hat a_{\fk{k}} + \hat a_{\fk{k}}^\dagger\right) 
\,,
\\
\delta\hat\varrho_{\fk{k}}
&=&
i\sqrt{\frac{\omega_{\fk{k}}}{V_Q}
\left(g_{\rm 2D}+\frac{\f{k}^2}{4m\varrho_0}\right)^{-1}}
\left(\hat a_{\fk{k}} - \hat a_{\fk{k}}^\dagger\right) 
\,,
\ea
with the well-known Bogoliubov dispersion relation
\bea
\label{dispersion}
\omega_{\fk{k}}^2=\mu\,\frac{\f{k}^2}{m}+\frac{\f{k}^4}{4m^2}
\,,
\ea
where $\mu=g_{\rm 2D}\varrho_0^{\rm 2D}$ is the (initial) chemical
potential. 

For a perfectly scaling condensate, the two-point function in
Eq.~(\ref{correlation}) yields the density-density correlations
at all times, and in particular also the frozen correlations
(the density contrast) after horizon crossing.
We can thus calculate the Fourier components
\bea
\label{spectrum_2d}
C_{\rm 2D}(\f{\kappa})
=
\int d^2\rho\frac{\langle\delta\hat\varrho(t,\f{0})
\delta\hat\varrho(t,\f{\rho})\rangle}{\varrho_0^2(t)}\,
e^{i\fk{\kappa}\cdot\fk{\rho}}
=\frac{g_{\rm 2D}|\f{\kappa}|}{\mu\sqrt{4m\mu+\f{\kappa}^2}}
\,.
\ea
in order to obtain the spectrum.

For large wavelengths, the spectrum is linear in the (co-moving)
momentum $\f{\kappa}$ in complete agreement to the curved space-time
analogy. 
For large (co-moving) momenta $\f{\kappa}$, where the dispersion
relation~(\ref{dispersion}) changes from linear to quadratic and the
curved space-time analogy breaks down, the spectrum of the two-point
correlation function approaches a constant value corresponding to a 
Dirac $\delta$-contribution 
$C(\f{\rho},\f{\rho}') \propto \delta(\f{\rho} - \f{\rho}')+\dots$. 
Interestingly, after subtracting this local term, we obtain a
scale-invariant spectrum $C(\f{\kappa}) \propto 1/\f{\kappa}^2$ for
large $\f{\kappa}$ -- in analogy to cosmic inflation, where the
spectrum is scale-invariant, too (in 3+1 dimensions it is $1/k^3$).
However, one should bear in mind that this scale-invariance occurs in
a region where the curved space-time analogy breaks down. 

For addressing the question of whether this effect (amplification of
the initial quantum fluctuations of the phon modes) can be measured,
we estimate the order of magnitude of the relative density-density
correlations $C(\f{\kappa})$ for $\f{\rho}\neq\f{\rho}'$.
Since the two-dimensional coupling constant, $g_{\rm{2D}}$, is
proportional to the ratio of the scattering length, $a_s$, and the
perpendicular extension of the condensate, $a_z$, we have typically 
\bea
\frac{\langle\delta\hat\varrho(\f{\rho})
\delta\hat\varrho(\f{\rho'})\rangle}{\varrho_0^2}
=\ord\left(\frac{a_s}{a_z}\right)
\,.
\ea
Note, however, that this can only be seen as a rough estimate of the
order of magnitude, since the two-point function is logarithmically
divergent at small distances (where eventually the employed
approximations break down).
Remembering the conditions~(\ref{conditions}) in 
Section~\ref{Dimensional Reduction}, we see that the relative density
contrast must be significantly smaller than one -- which was to be
expected.
However, the fluctuations could be on the percent-level 
(see Section~\ref{Summary}) and thus might well be measurable. 

\section{Density-Density Correlations in 3D}\label{3D}

For an expanding 3D condensate with quartic coupling the situation is
more complicated, since it does not show the perfect scaling behaviour
as in two dimensions. 
Therefore, it is not possible to infer the final (frozen) density 
(or phase) fluctuations from the initial state exactly as in the
previous case. 
Fortunately, for long wavelengths, we may exploit the effective
space-time analogy and derive the evolution using the tools known from
cosmology. 
To this end, we transform onto co-moving coordinates and thus
diagonalise the effective PGL-metric (\ref{PGL}). 
Note that, in three (spatial) dimensions, the background density
scales as $\rho(t)=\rho(0)/b^3(t)$.
Assuming the background density to be smooth (Thomas-Fermi limit), 
the evolution equation for the scale factor -- after switching off the 
trap -- reads, cf.~\cite{Kagan}
\bea
\ddot b(t) = \frac{\omega_0^2}{b^4(t)} 
\,.
\ea
Similar to the previous case, the scale factor accelerates 
(the interaction energy is transformed into kinetic energy) 
after switching off the trap, but quickly assumes a linear
behaviour, $b \propto t$ (for $t \gg 1/\omega_0$).

The following calculations are most conveniently performed using the
laboratory time but co-moving spatial coordinates.
One advantage of the analogy to curved space-times is the inherited
freedom of choosing suitable coordinates, where the independence of
the final result follows automatically from covariance.
The equation of motion of the field modes $\phi_{\fk{\kappa}}$ with 
co-moving wavenumber $\f{\kappa}$ assumes the form
\bea
\left( \frac{\partial^2}{\partial t^2}
        + 3 \frac{\dot b}{b} \frac{\partial}{\partial t}
        + \frac{ c_0^2 \f{\kappa}^2 }{ b^5 } \right)
                \phi_{\fk{\kappa}} = 0 \,.
\label{3d_eom}
\ea
Initially, for the trapped condensate, the damping term 
$3\dot\phi_{\fk{\kappa}}\dot b/b$ vanishes and the
modes oscillate freely.
During the expansion, the third term decreases whilst the second one
increases and eventually dominates the latter one -- in complete
analogy to an over-damped oscillator, the modes freeze in.
Because most phonon modes have frequencies much larger than the
trapping frequency (e.g., \cite{Ozeri}), this horizon crossing and
freezing process happens during the period of linear expansion 
$b(t) = \alpha t$ (with $\alpha \approx 0.82 \omega_0$).
In view of the adiabatic theorem, the initial vacuum state for these
(initially) rapidly oscillating modes is preserved (adiabatic vacuum)
during the initial non-linear expansion. 
Thus the equation of motion describing the freezing in of modes
can be simplified considerably by inserting $b(t) = \alpha t$ 
\bea
\left(
\frac{\partial^2}{\partial t^2}
        + 3 \frac{1}{t} \frac{\partial}{\partial t} +
        \frac{c_0^2 \f{\kappa}^2}{\alpha^5 t^5}
\right) \phi_{\fk{\kappa}} = 0
\ea
and solved analytically in terms of Bessel functions \cite{Abramowitz}
\bea
\phi_{\fk{\kappa}} 
= 
C_{\fk{\kappa}}^{(1)} \frac{1}{t} H_{2/3}^{(1)}
\left( \frac{2}{3} \frac{c_0 \kappa}{\alpha^{5/2}} t^{-3/2} \right)
+ 
C_{\fk{\kappa}}^{(2)} \frac{1}{t} H_{2/3}^{(2)}
\left( \frac{2}{3} \frac{c_0 \kappa}{\alpha^{5/2}} t^{-3/2} \right)
\,,
\ea
where $H_{2/3}^{(1,2)}$ denote the Hankel functions with the index
$\nu=2/3$ and $C_{\fk{\kappa}}^{(1,2)}$ are the corresponding
integration constants.
As one would expect, for early times, the modes 
$\phi_{\fk{\kappa}}$ oscillate $\dot\phi_{\fk{\kappa}}^{(1,2)}=
\pm i\omega_{\rm ad}\phi_{\fk{\kappa}}^{(1,2)}$.
It turns out that the Hankel functions $H_{2/3}^{(1,2)}$ have the
proper asymptotic behaviour for early times such that the 
integration constants $C_{\fk{\kappa}}^{(1,2)}$ can be replaced by
(time-independent) creation and annihilation operators 
(in analogy to the previous Section) associated to
the initial adiabatic vacuum state for the quantised phonon modes
$\hat\phi_{\fk{\kappa}}$ in the Heisenberg picture
\bea
C_{\fk{\kappa}}^{(1)} 
\rightarrow
\sqrt{\frac{\pi}{6}\frac{g}{\alpha^3}\frac{2}{V_{\rm Q}}}
\;\hat a_{\fk{\kappa}}
\,,
\ea
where the pre-factor can be derived by imposing the canonical
commutation relations for the conjugate variables $\phi$ and
$\delta\varrho = \dot \phi/g$ obtained from the Lagrangian~(\ref{L-phi})
for co-moving coordinates.
Note that the argument of the Hankel functions coincides (up to a constant)
with the proper time, $\tau\propto\frac{2}{3}\alpha^{-5/2}t^{-3/2}$, during 
the phase of linear expansion.

For late times $t \uparrow \infty$, the modes $\phi_{\fk{\kappa}}$
approach a constant value (due to horizon crossing and freezing) which
can be derived inserting the asymptotic behaviour of the Hankel
functions $H_{2/3}^{(1,2)}$ \cite{Abramowitz}.
The frozen late-time $t \uparrow \infty$ expectation value for
phase-phase correlations reads  
\bea
\langle\hat\phi_{\fk{\kappa}}^2\rangle 
=
\int d^3\rho\,\cos(\f{\kappa}\cdot\f{\rho})
\langle
\hat\phi(\f{\rho})\hat\phi(\f{0})\rangle
= 
\frac{g}{6\pi}\,
\left[\Gamma\left(2/3\right)\right]^2
\,
\frac{\alpha^{1/3} 3^{4/3}}{c_0^{4/3}}
\kappa^{-4/3}
\,.
\ea
Employing Eq.~(\ref{density}), we can calculate the density
fluctuations, $\delta\varrho = - \dot{\phi}/g$ 
(in co-moving coordinates, where $\vau_0$ vanishes), 
and obtain the spectrum of the relative density-density correlation 
function at late times 
\bea
C_{\rm 3D}(\f{\kappa})
=
\int d^3\rho 
\cos( \f{\kappa}\cdot\f{\rho} )
\frac{\langle\delta\hat\varrho(\f{\rho})\delta\hat\varrho(\f{0})\rangle}
{\varrho_0^2}
= 
\frac{[\Gamma(1/3)]^2 3^{2/3}}{6\pi}
\frac{\xi c_0^{1/3}}{\varrho_0\alpha^{1/3}}
\,\kappa^{4/3}
\,,
\label{fluc_3d}
\ea
where everything on the right-hand side (healing length etc.) refers
to the initial state. 
Interestingly, one obtains the same spectra ($k^{-4/3}$ for the phase
and $k^{+4/3}$ for the density fluctuations) for a condensate at rest  
after sweeping through the phase transition at $g=0$ by means of a
time-dependent coupling $g(t)$ \cite{Schuetz}.

In contrast to the quasi-2D case with perfect scaling, the above
results rely on the curved space-time analogy and thus are only valid
for wavelengths far above the healing length.
Furthermore, in three dimensions, the healing length scales like 
$\xi(t)=\xi(0)b^{3/2}(t)$ and grows faster 
(again in contrast to the quasi-2D case) than the wavelengths 
$\lambda(t)=\lambda(0)b(t)$ in laboratory coordinates.
In order to obtain a rough estimate of the maximum amplification
effect, we insert the maximum co-moving wavenumber
$\kappa = 1/\xi  ( \alpha /\omega_\xi)^{1/4}$
of the phonon modes whose
wavelength exceeds the healing length during the relevant period of
their evolutions  (i.e., until horizon-crossing and freezing)
\bea
\kappa^3C_{\rm 3D}(\kappa)
&=&
\frac{4\sqrt{\pi}[\Gamma(1/3)]^2(\tilde\alpha)^{3/4}}{3^{1/3}}
\,\sqrt{a_s^3\varrho_0}
\left(\frac{\omega_\xi}{\omega_0}\right)^{-3/4}
\nn
&\approx& 
30.3
\sqrt{a_s^3\varrho_0}
\left(\frac{\omega_\xi}{\omega_0}\right)^{-3/4}
\label{fluc_3d_2}
\,.
\ea
Here $\omega_\xi = c_0/\xi(0)$ denotes the (initial) frequency of the
phonons with the (co-moving) wavenumber $1/\xi(0)$ and 
$\alpha=\tilde\alpha\omega_0$ where $\tilde\alpha\approx0.82$ is a
dimensionless constant. 
In contrast to the quasi-two-dimensional case, the size of the
fluctuations now depends on the diluteness parameter 
$\sqrt{a_s^3\varrho_0}$ (instead of the ratio $a_s/a_z$),
which must be small for the employed approximations to apply 
$\sqrt{a_s^3\varrho_0}\ll1$.  
Furthermore, the ratio $\omega_\kappa/\omega_0 = c_0 \kappa/\omega_0$
should be large $\omega_\kappa/\omega_0\gg1$ if we want a sufficiently
long period of linear expansion $b(t) = \alpha t$.
However, the smallness of these parameters can partly be
compensated by the numerical pre-factor $30.3$ such that the final
effect can be on the percent-level (see the next Section). 

The spectrum and size of the frozen-in correlations $C_{\rm 3D}$ 
could be measured by obtaining a density map of a slice of the
condensate (tomographic imaging).
By making a projective (e.g., absorption) image instead, one would
average over the fluctuations in the (spatial) direction of projection.
While the spectrum remains unchanged, this averaging yields an
additional suppression by a factor of order $\xi/l_z$ with $l_z$
denoting the transversal (i.e., in direction of projection) 
extension of the condensate.

\section{Summary}\label{Summary}

The objective was to investigate and to exploit the analogy between
phonons in expanding Bose-Einstein condensates on the one hand and
quantum fields in an expanding universe on the other hand.
The advantages of this curved space-time analogy are two-fold:
Firstly, it facilitates the application of all the concepts, tools,
and effects known from general relativity (such as horizons) and
thereby fosters a better understanding of condensed-matter
phenomena in terms of universal geometrical concepts.  
Secondly, this analogy enables a theoretical and possibly experimental
investigation of exotic quantum effects known from cosmology.

After a discussion of the effectively lower-dimensional behaviour of
strongly constrained BECs in Section~\ref{Dimensional Reduction}, the
emergence of an  metric governing the propagation of low-energy
phonons is discussed in Section~\ref{Effective Geometry} 
for arbitrary power-law self-coupling in any dimension.
In complete analogy to cosmology, the introduction of co-moving
coordinates is advantageous for the description of many phenomena.
It turns out that the full field operator 
(in a harmonic trapping potential) possesses perfectly scaling
solutions exactly if the effective metric is flat in terms of the 
co-moving coordinates.
In a freely expanding Bose-Einstein condensate, an effective sonic
horizon is formed, i.e., two points at fixed (co-moving) spatial
positions whose distance exceeds the horizon size cannot be connected
anymore by (low-energy) phonons, cf.~Section~\ref{Horizon Analogues}.
The formation of this effective horizon implies the amplification of
the quantum fluctuations (horizon-crossing and freezing) already known
from cosmology, which has been calculated explicitely for quasi-2D and
3D condensates with quartic coupling in Sections~\ref{2D}
and~\ref{3D}.

In order to obtain an explicite estimate for the size of the derived
effect, let us consider a condensate of $10^5$ sodium atoms inside a
highly anisotropic trap with the trapping frequencies 
$\omega_\perp/2\pi = 790\,\rm{Hz}$ and
$\omega_\parallel/2\pi = 10\,\rm{Hz}$, cf.~\cite{Gorlitz}.
The thickness of the disk, $a_z = 0.746\,\mu\rm{m}$, is smaller than
the healing length, $\xi = 1.34\,\mu\rm{m}$, but still much larger than
the scattering length, $a_s = 2.8\,\rm{nm}$, complying with the
hierarchy of scales derived in Sec.~\ref{Dimensional Reduction}.
To estimate the order of magnitude of the density-density correlations,
we consider a mode with (co-moving) wavenumber $\kappa = 2\pi/\xi$.
For the relative density contrast inside a volume of size $\xi^2$ we
obtain $C(\kappa)/\xi^2 = 1.79\,\%$.
For the modes in the linear regime of the spectrum, horizon crossing
and thus freezing in occurs shortly after the trap is switched off.
Initially the horizon is at infinity, but very quickly settles at 
$r_{\rm{horizon}}(t\uparrow\infty) = 3.28\,\mu\rm{m}$.
Hence already after one $e$-fold, when the radius of the atom cloud has
increased by a factor $e\approx2.7$, all modes with wavelength larger
than $\xi$ are frozen in and the fluctuations are transformed into a
density contrast. 
With the correlation function depending on the ratio $a_s/a_z$, the
effect can be enhanced by confining the condensate more tightly in the
perpendicular direction.
However, if  $a_s/a_z$ is not small, the condensate is no longer
described by the usual Gross-Pitaevski\v\i\/ Lagrangian and our
analysis does not apply anymore in this form (e.g., the perfect
scaling breaks down).
For such a case, the (initial) correlation spectrum 
(with appropriate non-quartic coupling, cf.~\cite{Lee,Gies}) 
was calculated in Ref.~\cite{Gies} and differs from our result 
in Eq.~(\ref{spectrum_2d}). 

As an example for the 3D case, let us consider $^{87}$Rb atoms inside
a spherically symmetric trap.
For a condensate consisting of $10^7$ atoms and a trapping frequency
$\omega_0 / 2 \pi = 200\,\rm{Hz}$, which are potentially realisable
parameters, cf.~\cite{Streed}, the Thomas-Fermi radius is
$R = 12.2\,\mu\rm{m}$.
The minimal frequency of phonons
$\omega_{\rm{min}} = 2\pi c_s/R = 5582\,\rm{Hz}$ is well above the
trapping frequency $\omega_0$.
Hence the freezing of modes takes place during the period of linear 
expansion and the approximation $b(t) = \alpha t$ in the evolution
equation of the phonon modes is justified.
A rough estimate of the maximum effect according to
Eq.~(\ref{fluc_3d_2}) yields 
\bea
\frac{\langle\delta\hat\varrho(\f{\rho})
\delta\hat\varrho(\f{\rho'})\rangle}{\varrho_0^2}
=\ord\left(2\,\%\right)
\ea
i.e., a potentially measurable effect

So far, we have assumed zero temperature.
Of course, initial thermal fluctuations would -- provided that there
is no thermalisation during the expansion of the condensate -- also be
amplified by basically the same mechanism as the quantum
fluctuations. 
In order to ensure that the quantum fluctuations are larger than the 
(initial) thermal fluctuations, the temperature must be small enough
such that the thermal occupation of the phonon modes under
consideration is negligible.
For the parameters used above, the relevant temperature scales
are $T(1/\xi) = 0.1\,\rm{nK}$ for the Q2D and 
$T(\kappa) = 3.9\,\rm{nK}$ for the 3D condensate, respectively.
By increasing the particle number and/or the trapping frequency, the
requirements regarding the experimental temperature can be relieved.
(Another way to discriminate between thermal and quantum fluctuations
is the spectrum.)

In summary, expanding Bose-Einstein condensates facilitate the
experimental simulation of exotic effects of quantum fields in
curved space-times.
On the other hand, the amplification mechanism under consideration
sets the ultimate quantum limit of accuracy for time-of-flight
experiments: 
No matter how smooth the initial cloud can be prepared, the frozen and 
amplified quantum ground-state fluctuations generate noticeable density  
perturbations.

\ackn

The authors acknowledges valuable conversations with 
U.~R.~Fischer and W.~G.~Unruh.
This work was supported by the Emmy-Noether Programme of the 
German Research Foundation (DFG) under grant No.~SCHU 1557/1-1.  
Further support by the COSLAB programme of the ESF, 
the Pacific Institute of Theoretical Physics, 
and the programme EU-IHP ULTI is also gratefully acknowledged. 

\section*{References}

\end{document}